# SubwayPS: Towards Smartphone Positioning in Underground Public Transportation Systems

**Thomas Stockx**
Expertise Ctr. For Digital Media
Hasselt University - tUL - iMinds
thomas.stockx@student.uhasselt.be

**Brent Hecht**
GroupLens Research
University of Minnesota
bhecht@cs.umn.edu

**Johannes Schöning**
Expertise Ctr. For Digital Media
Hasselt University - tUL - iMinds
johannes.schoening@uhasselt.be

**ABSTRACT**
Thanks to rapid advances in technologies like GPS and Wi-Fi positioning, smartphone users are able to determine their location almost everywhere they go. This is not true, however, of people who are traveling in underground public transportation networks, one of the few types of high-traffic areas where smartphones do not have access to accurate position information. In this paper, we introduce the problem of underground transport positioning on smartphones and present *SubwayPS,* an accelerometer-based positioning technique that allows smartphones to determine their location substantially better than baseline approaches, even deep beneath city streets. We highlight several immediate applications of positioning in subway networks in domains ranging from mobile advertising to mobile maps and present *MetroNavigator,* a proof-of-concept smartphone and smartwatch app that notifies users of upcoming points-of-interest and alerts them when it is time to get ready to exit the train.

**Categories and Subject Descriptors**
H.5.m. Information interfaces and presentation (e.g., HCI): Miscellaneous.

**INTRODUCTION & MOTIVATION**
A smartphone's ability to detect its location is the cornerstone of mobile applications in domains ranging from location-based services to mobile crowdsourcing. While technologies like GPS and Wi-Fi positioning have made location detection possible in most places smartphone users go, this is not true everywhere. Underground public transportation systems, which are largely inaccessible to GPS signals and often out of range of Wi-Fi networks, represent a particularly important type of space in which smartphones cannot reliably determine their location. As a result, the millions of locals and tourists who ride subways around the world each day [21] cannot take full advantage of popular location-aware smartphone applications – including mobile map apps – while traveling to and from their destinations.

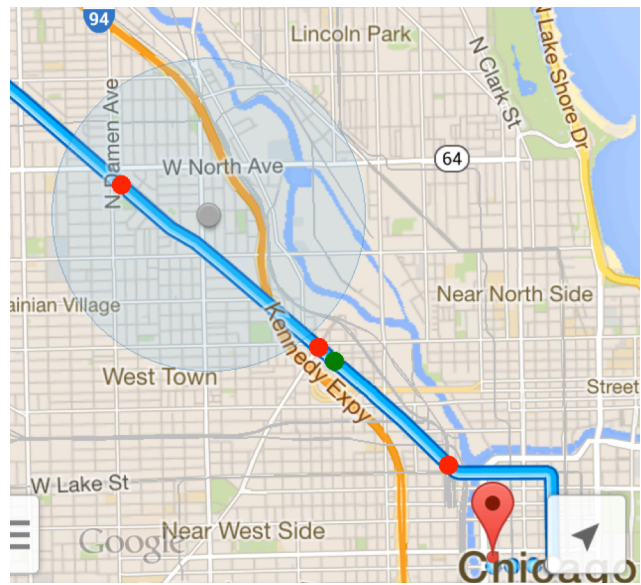

**Figure 1:** A screenshot of the Google Maps iPhone application used on the Blue Line subway in Chicago. The gray dot shows the smartphone's estimated position, while the green dot shows the actual position of the user's train. Red dots are underground stations. One could easily miss their stop if they relied solely on current positioning technology. *SubwayPS* works towards addressing this issue by providing improved positioning while traveling in underground public transportation networks (Base map © Google Maps 2014).

In this paper, we present *SubwayPS*, the first positioning technique that allows modern smartphones to determine their location while traveling in an underground public transportation network. Our technique, which we demonstrate can achieve reasonable – though not perfect – accuracy, does not depend on any instrumentation of the environment. Instead, SubwayPS merely requires an accelerometer and gyroscope and an input origin location and end destination. Accelerometers and gyroscopes are included in most recent smartphones. Origins and destinations can be easily extracted from routing requests (especially if this type of positioning technology is built into a mobile OS) or inferred from user behavior (as in context-aware technologies like Google Now [6]).

Smartphone positioning in underground public transportation networks affords wide-ranging benefits. The most basic such benefit is that SubwayPS simply puts subway riders "back on the map". Figure 1 shows the result of using Google Maps for iOS while traveling through an underground portion of the Chicago Blue Line. Because the app cannot obtain position information, it is rendered effectively useless for navigation and orientation purposes.

Our technique can make mobile maps as useful in underground transportation networks as they are above ground. For example, a SubwayPS-enabled navigation app could much more effectively help tourists navigate foreign subway systems, especially when they do not speak and cannot read the native language. Such an app could, for instance, inform tourists who cannot understand the announcements of when to prepare to exit the train, an especially important concern when the train is crowded.

SubwayPS can also enable location-based applications beyond basic navigation and orientation. For instance, if our technique were built into the location API of a mobile OS, smartphone users would be able to use cached content from apps like Yelp to easily see the restaurants that would be available if they were to get off at the next stop. In addition, mobile advertising becomes possible underground (e.g. "Exit at the next stop to get to restaurant X and receive $5 off").

The core contribution of this paper is our underground public transportation positioning technique SubwayPS, along with three evaluations that show that our technique can achieve reasonable accuracy and substantially outperforms baseline approaches on multiple subway systems. Below, we first put this core contribution in the context of related work. Next, we present SubwayPS in detail and introduce *MetroNavigator*, a proof-of-concept application based on SubwayPS. We then demonstrate the accuracy of SubwayPS using log data and in-the-wild user studies, showing that it performs substantially better than baseline approaches. We also highlight qualitative feedback from participants, which included *"I want to have it integrated in Google Maps"* and *"I feel safer when I can see that we have stopped close to the next station when we're stopped in the middle of a tunnel"* (P11). Finally, we close by highlighting in more detail the many applications of subway positioning and discussing several means by which accuracy can be improved in future work.

**RELATED WORK**

The work presented here is informed by research from three areas: inertial navigation, transportation mode detection, and other approaches to positioning in public transportation networks. Below, we discuss each of these areas in turn.

**Inertial navigation**

SubwayPS uses an approach to positioning motivated by the literature on accelerometer- and gyroscope-based inertial navigation. Although inertial navigation approaches have not been adapted to the context of public transportation positioning as they are here, they have been shown to be successful for other purposes, especially pedestrian indoor navigation (e.g. [1,5,13,14]). For instance, Robertson et al. [13] presented a simultaneous localization and mapping (SLAM) approach for pedestrians using only foot-mounted inertial sensors. Their method, *FootSLAM*, can be used to generate an approximate 2D map of an indoor layout with just 10 minutes of walking.

Other accelerometer-based approaches have combined inertial navigation with GPS and other positioning signals for improved accuracy. The *NavMote Experience* [4], for example, includes a dead reckoning module, self-organizing wireless ad hoc networks and an information center with a map database. Another example is the *Embedded GPS/RFID/Self-contained Sensor System* by Kourogi et al. [9]. As described in the name of their system, Kourogi and colleagues integrated self-contained dead reckoning sensors (accelerometers, gyrosensors and magnetometers) with GPS and an RFID tag system to adjust for errors in position and direction that occur in the dead reckoning process.

A number of inertial navigation-based positioning techniques for indoor navigation and related purposes have been developed for smartphones. These techniques generally use the detection of walking steps ("step detection") together with a method for heading determination. Such techniques have seen huge accuracy improvements in recent research (e.g. [8,10,11,15,16,17]) but are unsuitable for localization in underground public transport systems where step detection is not an option.

**Transportation Mode Detection**

Wang et al. [19] showed that accelerometers on mobile devices can be used to detect a user's mode of transport (e.g. walking vs. biking vs. driving). In this research, they used a formula for acceleration synthesization to learn the patterns of different transportation modes. We implement their formula in our SubwayPS technique by using it to detect if a train is moving or not. Hemminki et al. [7] have recently achieved improved accuracy on transport mode detection by providing a new algorithm for estimating the gravity component and key characteristic pattern detection, which involves the detection of mode- specific patterns such as step detection for walking and smooth accelerations for vehicle movements.

**Positioning in Public Transportation Systems**

The final area of work related to SubwayPS comes from the literature on positioning in public transportation systems. For instance, the EasyTracker system [2] provides a location tracker for public transport that uses GPS traces from above-ground transit vehicles (busses), but requires all vehicles to be instrumented with equipment such as GPS receivers and inertial navigation sensors. Zhou et al. [20] developed a positioning technique that combines a variety of sensor measurements from mobile devices owned by the people on an above-ground transit vehicle to calculate an average location. Thiagarajan et al. [18] describe a crowd-sourced transit tracking system intended to improve real-time arrival and departure estimates. As part of this research, Thiagarajan and colleagues looked at subway systems, but their approach is not designed for positioning on mobile devices and requires a connection to a tracking server, which is often not available while on a subway.

To summarize, in this paper we present the first smartphone positioning technique (SubwayPS) for underground public transportation systems that relies solely on the user's mobile device. As a result, our technique does not require the infrastructure [2], connectivity [18], or the critical mass of users [20] needed by existing approaches in this area. Instead, SubwayPS merely requires a single accelerometer and gyroscope, both of which are built into most modern smartphones. In addition, SubwayPS has important privacy benefits: because all calculations are done locally on the user's mobile device, SubwayPS does not share a user's position with any central server or third parties.

**THE SUBWAYPS TECHNIQUE**

With the SubwayPS technique, we show that it is possible to achieve reasonable positioning accuracy on underground public transportation networks. Our results suggest that SubwayPS is usable as-is on most subway systems in the world. We also show how accuracy can be improved by fine-tuning certain parameters to the characteristics of a specific underground public transport system.

SubwayPS makes use of a smartphone's accelerometer and gyroscope by combining the accelerometer measurements on all three axes to detect underground train movement. The gyroscope is used to filter the gravity factor out of the accelerometer measurements. As such, the accelerometer reads (x = 0 m/s², y = 0 m/s², z= 0 m/s²) when the mobile device is at rest and the coordinate system is defined relative to a default orientation of the accelerometer sensor. The "axes are not swapped when the device's screen orientation changes" [22]. In our implementation of SubwayPS, we made use of the virtual linear accelerometer provided by the Android operating system, which handles the gravity removal internally.

Collecting accelerometer data in a pretest on multiple underground public transport systems showed that station locations overlapped with periods in our data collection when there was generally less acceleration *on all axes of the accelerometer*. This is due to the fact that train movement results in multiple small accelerations in all directions (i.e. the small amount of shaking back and forth experienced by passengers when a train is at high speed). As such, by measuring the amount of variance on all axes, it is possible to detect whether the train is moving or not.

Figure 2, which shows data collected on the London Underground subway system, demonstrates this phenomenon in more detail. Figure 2a depicts the accelerations measured on each axis of the accelerometer over a small period of time. Stations are clearly visible in the data in that they have accelerations closer to zero and less variation (around minute 1 and 2.7 in Figure 2a). By calculating a general acceleration value based on the three-axis input of the linear accelerometer data, it is possible to derive a statistic *a* that has a large value if there is a large amount of acceleration on any (or all) of the axes. To calculate this value, we use the following formula for acceleration synthesization [19]:

$$a = \sqrt{x^2 + y^2 + z^2}$$

The resulting *a* values are smoothed by a rolling average with a window of *n* samples. We found *n* = 100 samples at 50 Hz to be an appropriate general window. Lower values would confuse user movement of the smartphone as false train movement (see Study 2 for more details), while higher values would increase the delay between events (e.g. subway stops) and their detection by SubwayPS. Plotting out *a* results in a graph like that shown in Figure 2b.

By comparing the smoothed data with a certain threshold, it is possible to conclude if a train is moving or not. The *SubwayPS* technique uses multiple parameters to parse the data and uses states to define if a train is moving or not,

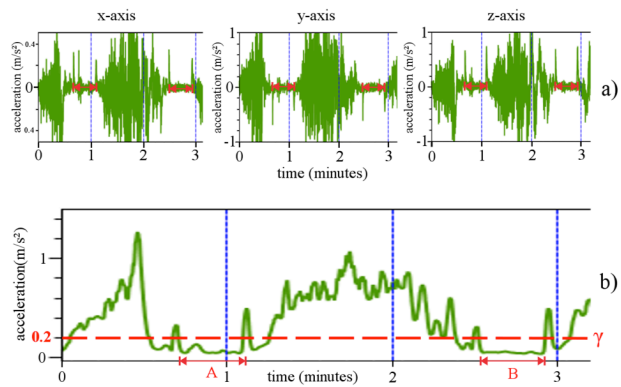

Figure 2: In Figure 2a, a visualization of accelerometer values measured on each axis is shown. The blue dotted lines indicate time in minutes while the red intervals indicate stations. The data was collected on the Central line in London. Figure 2b shows the result of the acceleration synthesization used in SubwayPS, smoothed by a moving average. This result is compared to the threshold to detect train movement.

based on a window of samples. The parameters considered are:

$n$ = *window size of moving average*

$\gamma$ = *threshold*

$\delta_{below}$ = min. amount of samples below threshold

$\delta_{above}$ = min. amount of samples above threshold

The values for each of the four parameters was determined empirically using data from four different subway systems as described below in the evaluation section. SubwayPS requires at least $\delta_{below}$ samples below the threshold before it marks the current accelerometer measurements as a stop, while at least $\delta_{above}$ samples above the threshold are required for SubwayPS to conclude that the train is moving. These values are used to prevent flagging user movement as train movement (Study 3).

The accuracy of SubwayPS on individual public transportation networks can be improved by customizing the values of $\delta_{above}$ and $\delta_{below}$ for each network. For instance, increased accuracy for the London and Cologne subways can be achieved by using the values in the second and third columns of Table 1, respectively. Trains in London accelerated more slowly, resulting in a longer period required to reach $\delta_{above}$, while in Cologne, trains accelerated rather quickly. Because the threshold was reached much faster in Cologne, $\delta_{above}$ could be increased to decrease the chance of false positives.

While the "world-wide" parameters were empirically determined to work reasonably well on all subway systems tested (see below), the parameters specifically tuned to the data we collected on each subway system can increase local performance. For example, using the world-wide parameters for SubwayPS on data collected in London results in an accuracy of 74.2%, while using the London parameters increases the accuracy percentage to 85% (Study 1).

Our stop detection method is complemented by a linear interpolation of the current position based on the time intervals between stations as defined in official timetables. By interpolating over time, we can estimate the absolute location between the previous and the next station. If SubwayPS detects a stop that is not scheduled, SubwayPS marks it as an "in-between" stop and uses the interpolation of location over estimated time to show a position to the user. Using this method we can provide the users with a position estimate of where exactly the train stopped between two stations. We consider any stop that occurs with less than 70% of the estimated time between stations to be an "in-between" stop, a number that was determined empirically from our Study 1 data (see below). This percentage can be set to a higher value if a timetable is available with seconds-level precision, which was not the case with the four subway systems considered here, all of which list timetables at the precision of minutes.

## IMPLEMENTATION

We implemented a working version of SubwayPS on the Android smartphone platform, as well as a proof-of-concept SubwayPS-based Android application called MetroNavigator, which has both smartphone and smartwatch versions. We first describe our implementation of core SubwayPS and then discuss MetroNavigator.

### Implementation of SubwayPS

SubwayPS makes use of the virtual linear accelerometer class provided by the Android OS. It "measures the acceleration force in m/s$^2$ that is applied to a device on all three physical axes (x, y, and z)" [22] without the influence of gravity. These accelerometer measurements are interpreted by the SubwayPS technique, which is implemented as a background service and sends *intents* (messages) such as *StationDetected* or *MovingDetected* to the Android OS, which then forwards them to all applications that implement a listener for these intents. Any application (e.g. MetroNavigator) could implement an intent listener and make use of our detection algorithm. This background service is just a proof of concept. Preferably, the *SubwayPS* technique would be implemented on an OS level such as within the location services of Android.

### MetroNavigator Smartphone Application

MetroNavigator (Figure 3) is a proof-of-concept application we built that uses SubwayPS as its positioning technology (by implementing an intent listener). The application's UI consists of two main parts. The upper part contains a miniature map with an indicator of the user's current position between the previous and next station, as well as the origin and destination station, the estimated time of arrival and the number of stops to the end station. The lower part of MetroNavigator's UI is used to display "event cards". Currently our *MetroNavigator* application supports multiple different types of event cards:

|  | World-wide | London | Cologne |
|---|---|---|---|
| $\gamma$ | 0.2 m/s² | 0.2 m/s² | 0.2 m/s² |
| $\delta_{below}$ | 250 samples | 250 samples | 250 samples |
| $\delta_{above}$ | 350 samples | 250 samples | 500 samples |
| $n$ | 100 samples at 50 Hz | 100 samples at 50 Hz | 100 samples at 50 Hz |

**Table 1: General, "world wide" parameters for SubwayPS, as well as those specifically tuned for London and Cologne.**

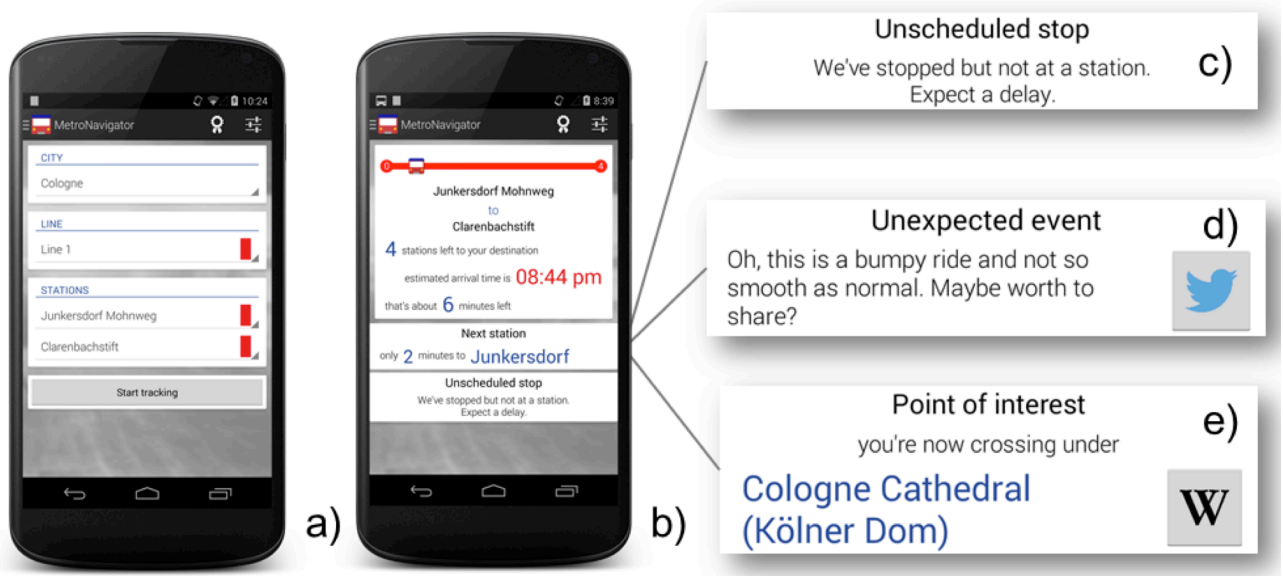

**Figure 3: MetroNavigator application showing different "event cards".**

1. The first, and most-often-used type is the notification event card shown in Figure 3b. When the train approaches the next station, it shows the name of the next station and can indicate possible connections. When the train approaches the final station, this is prominently indicated.
2. The application also triggers event cards when the train undergoes unplanned stops in between stations. This is shown in Figure 3c.
3. Another event card type notifies the user of "rude" driving by the train conductor based on measured accelerations. This is shown in figure 4 and one can share this event on Twitter (once connectivity is obtained) with a direct mention of the public transportation company.
4. *MetroNavigator* also contains point-of-interest (POI) event cards that contain information about POIs that are located above the user or at the next station (see Figure 3e) and provide a link to more information about the POI on Wikipedia, which is locally stored on the device.

**The MetroNavigator Smartwatch Extension**

We also implemented a smartwatch version of MetroNavigator that can be linked to an Android device running SubwayPS via Bluetooth. Due to screen size constraints, the smartwatch version has a much simpler UI than the smartphone application. It displays to users only the current or next station, as well as the estimated time of arrival. A screenshot of the smartwatch application is shown in Figure 5.

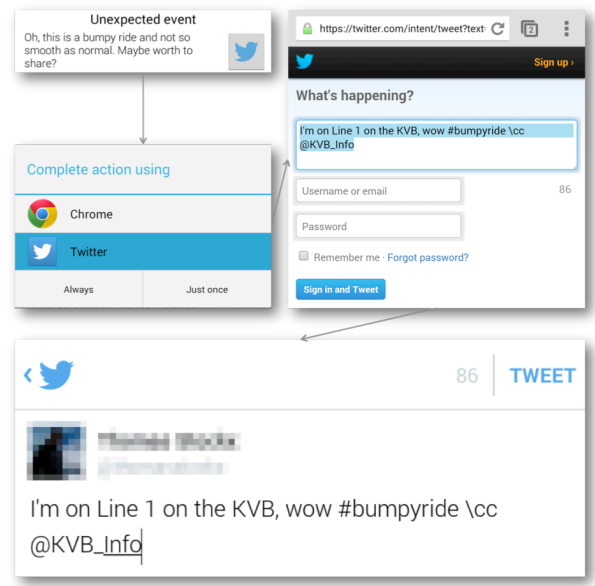

**Figure 4: Sharing an event card of MetroNavigator with a few clicks.**

The smartwatch version also has a feature not yet available in the smartphone version: when users reach the final subway station of their trip, the smartwatch version tells users in which direction to go to reach their destination (Figure 6). In addition, when selecting a POI in the smartphone app, the smartwatch will show the direction to go to reach the POI once the user has reach the next station.

**EVALUATION**

To evaluate SubwayPS, we conducted three separate studies with the help of the MetroNavigator application. The first study was focused on assessing the technical soundness of the positioning technique in a variety of public transportation systems around the globe using log data. The second study was designed to compare SubwayPS against a timetable-only baseline and to evaluate the effect of arbitrary smartphone movements by users. The third evaluation was focused on understanding the accuracy and user experience of the MetroNavigator application in the context of actual subway journeys made by locals and tourists.

**Study 1**

The goal of our first study was to collect accelerometer data from a variety of diverse subway systems and use this data to inform and evaluate SubwayPS.

*Participants & Apparatus*

For the data collection process, we developed an Android application that captures and stores the data measured by smartphone accelerometer sensors. The data collection application supports the public underground transportation network of four major cities, namely Brussels, London, Cologne, and Minneapolis. Each of these subway systems has unique characteristics: they use different trains, some go above ground, and others merge with vehicular traffic. We wanted to study a diverse set of subway systems in order to support broader generalizability.

To record data, participants first selected a city, a line and an origin and destination station. Participants were advised to select the start and end station of their journey before boarding the train and were told to mainly place their phone in their pockets.

We recruited twelve participants that downloaded the data collection application to their mobile device and asked them to record data "whenever they use an underground public transport system". We also asked them to record very long tracking segments to get tracks that are longer than normal underground rides so our algorithm could be tested exhaustively (21.51 minutes trip length vs. about 9 minutes in the follow up studies). Most of the participants were Android developers and computer science students and collected the data on a voluntary basis on their daily commutes or on business trips. These participants collected a total of about 70 tracks in a period of around two months and we received tracks from every underground transport system supported by the application.

*Results*

The average number of stations per track was 9.23 and the average track length was 21.51 minutes. We analyzed this data to determine the values for each of the SubwayPS parameters described above (e.g. $\delta_{below}$, $\delta_{above}$). Using the subway system-specific values shown in Table 1, we found that for London, 103 out of 120 stops (105 stations and 15 "in-between" stops) were classified correctly (85.8% excluding the start stations). 10 stations and 7 "in-between" stops were missed. No false positives (detection of a stop where there was none) were recorded.

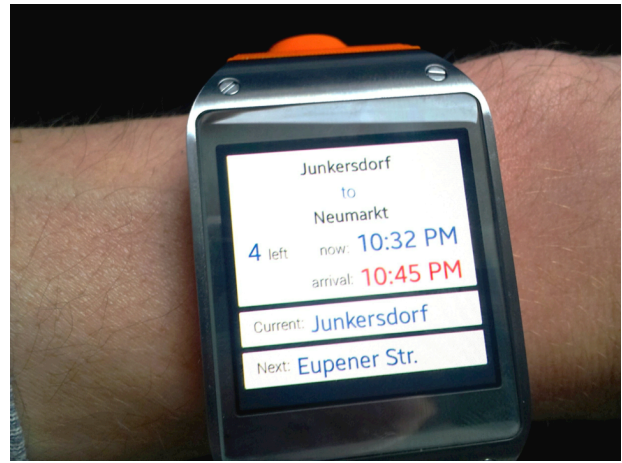

**Figure 5: Screenshot of the SubwayPS smartwatch application that provides basic feedback to the user without needing to take the smartphone out of their pocket.**

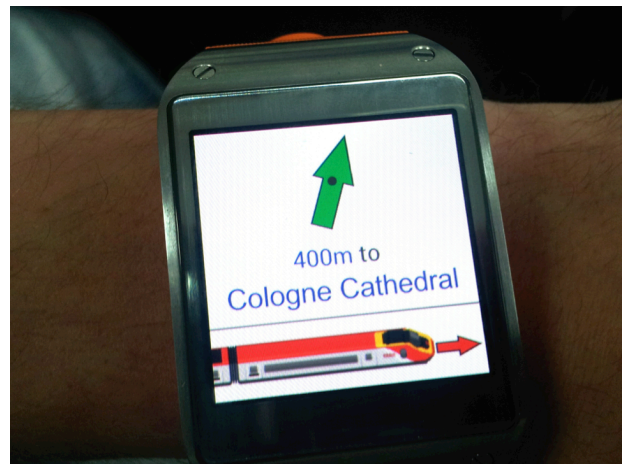

**Figure 6: Screenshot of the SubwayPS smartwatch application that provides guidance on in which direction to exit the final station to reach a certain POI.**

This can be directly compared to the results of the same tracks tested with the "world-wide" parameters, which caused SubwayPS to correctly detect and classify about 74.2% of these stops.

Similar results (around 85% in comparison to around 75% for "world-wide" parameters) were measured in the other three subway systems.

These results are somewhat comparable to those of a study done by Thiagarajan et al. [18]. Using four tracks in the subway system of the Chicago public transportation network (as opposed to our 70 tracks across 4 subway systems), Thiagarajan et al. were able to achieve around 55% accuracy (compared to our 85.8%), although it is important to note that their focus was on station detection rather than location prediction (which involves detecting in-between stops). In Study 3, we show that SubwayPS also strongly outperforms a "timetable" baseline.

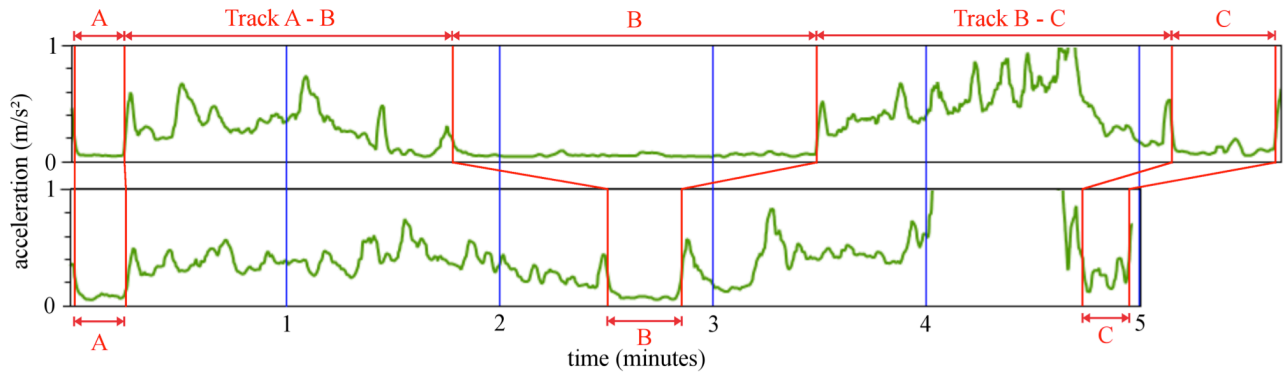

Figure 7: Example of day-to-day variability measured on the Piccadilly line in London. Both graphs picture the data calculated by *SubwayPS*. Stations A, B, and C are clearly visible. These graphs provide an example of how much variability there is in the measurements. Times between the same two stations can vary greatly (see Track A-B) with a difference of almost one minute. The period during which a train stands still at a station can also vary greatly (see B).

Examining where SubwayPS failed, we noticed that trains can run "smoother" on certain lines (such as parts of the Hammersmith & City line in London) than others in a given subway network. Because SubwayPS's parameters were tuned on a subway system-wide basis, this variation resulted in errors. We are working to implement track-level parameters, which should increase accuracy by a substantial margin.

Another issue arose out of huge variations in travel time for trips between the same two stations. We recorded multiple readings over multiple days for some lines and noticed a standard deviation of about 7.12 minutes for a trip that normally takes 29 minutes. We looked into this particular case, and found out it was due to extensive waiting times at stations when trains were overcrowded. These delays could cause misclassification due to the integrated time-tables, as "in-between" stops could be classified as a station if the trip takes too long. An example of these variations in trip length is visible in Figure 7. The graphs depict two trips between the same stations and differ greatly in the in-between measurements.

Finally, another source of error was the quality of accelerometer data. For example, badly calibrated gyroscopes in all data captured by Samsung devices resulted in incorrect measurements by the virtual linear accelerometer (e.g. accelerations of above 1 m/s² on some axes while at rest). It would be straightforward to correct for these errors on a device-by-device basis if these errors are known a-priori.

### Study 2

A second user study was conducted with the MetroNavigator application to directly compare the use of SubwayPS against an approach that merely interpolated from an official timetable.

*Procedure*

The study was conducted across two consecutive days between 8 and 6 pm in May 2014 on the Cologne, Germany subway. One experimenter engaged in 50 trips in which he randomly boarded a train, travelled with it for exactly seven stations and then left the train and boarded the next train arriving at that station. In order to compare against a timetable baseline, the experimenter recorded the exact time at which the train he was riding arrived at each station. No restrictions were placed on the placement of the smartphone in this study. The Study 2 experimenter performed common activities normally executed on a smartphone as reported by Böhmer et al. [3], which involved periodically having the phone in his hands and storing the phone in his pocket during the trips.

*Results*

Across all 50 trips, 282 of 335 stops (300 stations and 35 "in-between" stops) were classified correctly (82.7%, excluding the start stations; 85.0% including the start stations). 35 stations and 18 "in-between" stops were missed. Again, no false positive stops were recorded. These results are in line with our previous results from Study 1 and Study 3 presented later in this paper.

To evaluate SubwayPS against the timetable baseline, we compared the actual arrival time at each stop on all 50 trips to (1) the arrival time indicated by SubwayPS and (2) the arrival time indicated by the official subway schedule. Using a 30-second "tolerance" window, we found that SubwayPS tracked 39 entire trips with perfect accuracy (78% of trips), while using the official timetable, only 21 entire trips (42%) were accurate.

The challenges presented to timetables from the "ripple effects" of a single delay have led some public transport systems to abandon timetables for an "interval"-based approach. In this approach, which was recently examined by Pritchard et al. [12], service is guaranteed every *n* minutes rather than at specific times. To understand SubwayPS's performance in this type of subway network, we also evaluated SubwayPS against a relative time baseline, which uses just the reported travel time between stations. This baseline was able to track 25 out of 50 trips correctly (50%), as opposed to SubwayPS's 39 of 50.

The trip-level accuracy of SubwayPS is directly influenced by trip length and accumulation errors (if one station is missed, the trip is not tracked correctly), which are known issues for inertial navigation techniques (i.e. "drift" errors). Fortunately, there has been extensive research on solutions for recalibration in inertial navigation. We are currently working to adapt one of these solutions to a public transportation network context.

**Study 3**

To test the "in the wild" feasibility and appeal of SubwayPS and its proof-of-concept implementation in MetroNavigator, a user study was conducted. *MetroNavigator* was pre-installed on a Google Nexus 4 device and the device was handed out to randomly selected passengers waiting at different stations in the Cologne, Germany subway system (*Kölner U-Bahn*). Due to novelty effect concerns, we focused on the smartphone version of *MetroNavigator* for the evaluation (instead of the smartwatch version) and restricted our participant population to subway riders who were already using an Android device.

Sixteen participants (8 male, 8 female) with an average age of 39 years took part in the study. Eight participants were employees at different local companies in Cologne, seven participants were tourists, and one participant was a student at a local university. The study was conducted across two days between 8 and 10am (peak travel time) within the period of a week in February 2014.

*Procedure*

The experimenter approached passengers that were using Android devices at one of the selected subway stations to invite them to participate in the study. If they agreed, the experimenter read a quick script that explained the purpose and functionality of MetroNavigator and informed the participant that the experimenter would be riding with her/him to their final destination. Participants were then asked to enter their start and end subway stops and were also requested to "think aloud" when interacting with the application. The script was kept short as we did not want to cause our participants to miss their train. The experimenter boarded the train with the participants and recorded their reaction during the ride.

Due to the relatively short nature of subway trips and our desire to have participants engage with the application, participants were asked to hold the device in their hands during the trip and not store it in their pockets. The robustness of SubwayPS to normal user movement was established in Study 2. After leaving the train, participants filled out a background questionnaire about their age, gender and occupation, and the experimenter conducted a semi-structured interview. The experimenter also explained how the system derives position information by using the accelerometer of the smartphone and the smartwatch extension was explained to the participants to get feedback on its possible uses.

*Results – Accuracy*

The participants took the train for an average of 6 stations (including the start and end station) and the average time on the train was about 8.75 minutes. 78 out of 91 stops (80 stations and 11 "in-between" stops) were classified correctly (85.7 % excluding the start stations; 87.9 % also including the start stations). Eight stations and 5 "in-between" stops were missed. No false positives were recorded. Out of the 16 trips, 12 were tracked completely correctly (75%).

*Results – Qualitative Feedback*

During the interviews, 15 participants expressed a positive opinion of SubwayPS and were interested in using MetroNavigator (and other applications using its SubwayPS engine) in the future. We received comments like "*As my GPS does not work here, this is like a GPS for undergrounds trains – I want to have it integrated in Google Maps*" (P3) or "*It is just cool to see the train moving and stopping on the map – I feel safer when I can see that we have stopped close to the next station when we're stopped in the middle of a tunnel*" (P11). The main advantage of MetroNavigator for most participants was the ability to be guided to their destination stop (14 out of 16 participants) and general location awareness (12 out of 16 participants). "*Wow - This is like magic*" (P3) was the comment of one participant when the visualization of her current position stopped between two stations as the train did. She commented "*… at first, I though this app is a bit boring as it just moves a metro along with the schedule from station to station, but now I see, that it has GPS*".

Participants also enjoyed specific features of the MetroNavigator app. In particular, several of the tourists found value in MetroNavigator's POI information cards "*as [they] can easily build local knowledge of a city*" (P2). The smartwatch was generally thought to be a useful extension for such an application. One participant remarked "*as I do not want look at my smartphone all the time, this (SubwayPS) could be the killer app for smartwatches in the future*" (P13, an IT consultant).

During the ride of one participant the MetroNavigator application missed three stops in a row. She commented on that by saying "*Oops, it seems that the system missed a stop – that is not good. I would also be happy to help the system to detect the stops, if the system could help me to get out at the right station. That would be totally fine with me*" (P3, a tourist). Other critical comments were targeted at the limited set of POIs in the app (e.g. "*Can it also show shops and café places?*", P3) . In our prototype, we stored just a few POIs and the dataset could be easily extended.

## DISCUSSION AND CONCLUSION

In this paper, we presented SubwayPS, a smartphone positioning system that puts users "back on the map" when they are traveling in underground public transportation networks. SubwayPS does not require any instrumentation of the environment, meaning it can be implemented without any expense to the often-cash-strapped operators of public transportation systems. SubwayPS merely requires an accelerometer and a gyroscope – both of which are standard on many modern smartphones – and a start and an end destination as input, which can be inferred from user behavior (as in Google Now) or extracted from routing requests.

Our evaluation showed that SubwayPS works "out of the box" with four subway systems from two different continents. However, we also showed that SubwayPS's accuracy can be increased if its four parameters are tuned specifically for an individual subway network. This tuning is simple, and merely requires a single user to collect accelerometer data as they travel on the subway.

In the short term, SubwayPS is straightforward enough to be implemented directly into a mobile app as we have done with MetroNavigator, our proof-of-concept SubwayPS application. We have taken care to ensure that any developer who wishes to implement SubwayPS can do so by following the instructions (and using the parameters) laid out in the "SubwayPS Technique" section above. However, if SubwayPS (or a technology like it) were built into a mobile OS, it would have the greatest impact, allowing all location-aware mobile apps – including mobile map apps – to function while their users traveled in subway networks.

That said, before OS integration can occur, several limitations of SubwayPS must be addressed. In its current implementation, accuracy at a trip level is highly dependent on trip length due to accumulation errors. Fortunately, there are various solutions to this issue. For instance, in many subway networks, subway stations have localized Wi-Fi coverage. This means that traditional Wi-Fi positioning techniques can be used to correct any errors before they accumulate. In addition, per-stop accuracy can likely be improved by using relatively straightforward machine learning approaches rather than the simple, simple subway system-wide approaches considered here.

Future work is proceeding along three directions. First, we are working to implement Wi-Fi-based positioning correction directly into the SubwayPS technique to correct for drift errors. This should greatly increase per-trip accuracy rates. Second, we are working on developing more sophisticated stop detection approaches using trained models as well as making small changes to support per-track parameters. Finally, we are considered the means by which crowdsourcing may be used to collect training data, as suggested by P3 in Study 2.

## ACKNOWLEDGMENTS
This research was supported in part by a Google Faculty Research Award, a 3M Non-Tenured Faculty Award, a Yahoo! ACE Award, and BOF project R-5209. The authors would like to thank our anonymous reviewers for their feedback.

*Note: This version of the paper contains a fix for a reference issue that appeared in the original version.*